# Construction method for general phenomenological RANS turbulence model

Shuming Zhang[1,2], Haiwang Li[1,2], Ruquan You[1,2,*], Tinglin Kong[1,2], Zhi Tao[1,2,3]

[1] Research Institute of Aero-Engine, Beihang University Beijing, 100191, China
[2] National Key Laboratory of Science and Technology on Aero Engines Aero-thermodynamics, Beihang University Beijing, 100191, China
[3] School of Energy and Power Engineering, Beihang University Beijing, 100191, China
*Corresponding Author: youruquan10353@buaa.edu.cn

## ABSTRACT

This paper proposes a phenomenological Reynolds Averaged Navier-Stokes (RANS) calculation model based on physical constraints. In this model part of the source terms in the $\varepsilon$ equation was replaced with the deep learning model, using the standard $k - \varepsilon$ model as a template. The simulation results of this model achieved a high error reduction of 51.7 % compared to the standard $k - \varepsilon$ model. To improve the adaptability and accuracy compared to the convergence of the abnormal flow regime, the coordinate technology proposed in this study was used in the modelling process. For the training data, the $k-$ field and $\varepsilon-$ field were automatically corrected using this approach when the flow state deviated from the theoretical assumption. Based on the coordinate technology, a deep learning model for the source term of the equation was built, and the simulation error was reduced by 6.2 % compared to the uncoordinated one. From the results, the proposed coordinate technology can effectively be adapted to the underdeveloped flow state and assist in the more accurately modelling of the phenomenological RANS calculation model under a complex flow state.

## I. BACKGROUND

### A. phenomenological turbulence model construction

The traditional method of modelling turbulence involves performing a theoretical analysis and then modifying it according to the results of experiment [1–3]. In recent years, evolving data-driven machine learning methods have proved that the idea of using phenomenological modelling to replace theoretical modelling is completely feasible in engineering. Haghiri [4] completed heat flux closures in large eddy simulations (LES) using gene expression programming (GEP) machine learning technology, which significantly improved the accuracy of the prediction of wall heat transfer; Beetham et al. [5] proposed an RANS regression framework based on sparse regression technology, which improves the interpretability of machine learning methods for the RANS closed problem. Among many machine learning models, the deep learning [6] model as a black box has a universal approximation ability [7] and can express highly nonlinear flow field characteristics; thus, it has been widely used by authors such as Yarlanki et al. [8] who completed the correction of empirical parameters for the $k - \varepsilon$ model [9] with data-driven ideas rather than traditional approaches. Gamahara et al. [10] obtained the subgrid stress with the help of an artificial neural network, and the actual performance was found to be better than that obtained by traditional theoretical analysis. Srinivasan et al. [11] evaluated the performance of the flow-field improvements demonstrated by different depth learning frameworks. For the $k - \varepsilon$ model, Ling [12] proposed a new architecture based on the deep learning model, which can effectively capture the Galileo invariance in the process of anisotropic Reynolds stress prediction and showed that deep learning can improve the prediction accuracy of the turbulence model. Some of Ling's work also shows that it is feasible to embed deep learning technology into the computational fluid dynamics (CFD) solution in the early stages [13]. In addition, for similar purposes, Frezat [14] proposed a new modulus strategy with invariant characteristics based on a neural network.

The anisotropy of the eddy viscosity field poses challenges to the application of deep learning for RANS modelling. If the isotropic eddy viscosity model is used, the actual anisotropy of the eddy viscosity field is ignored. If anisotropy is considered, the convergence of the computation may get worse[15], which is not good for practical engineering applications. When there is an energy inverse cascade [16–18] in the flow field, the obtained eddy viscosity field may still be locally negative, resulting in the divergence of the CFD iteration. These problems make it difficult to build a turbulence model by deep learning which is trained by inverse calculating $\mu_t$. To solve this problem, the existing methods is combining theory and several assumptions to obtain the data for deep learning. The current research indicates that different assumptions introduced in the obtaining of the training data, such as the use of different formula to derive the $\varepsilon$ field[19–21], lead to different models.

When using the models to do CFD computation will lead to different prediction, which will cause the input gradually deviate from the situation of training data with the iteration of CFD [22]. The deviation will affect the prediction of neural network. Wu et al. [23] used statistical methods to estimate the uncertainty of prediction brought by the deviation of input. In addition, the focus on state of flow and areas may decrease the generality of the deep learning model, which needs to be avoided in the study[24]. The above research shows that the current RANS turbulence modelling is not an definite process, and the goal and results of modelling are not constant but vary by the assumption introduced into the model. Using this type of method will always introduce some uncertainty.

The data processing based on the assumption which is introduced manually will bring some uncertainty, and deep learning turbulence modelling can solve this issue. For example, Guo et al. [25] directly modelled the average flow field using a convolutional neural network [26], directly generating the average flow field from the boundary conditions. Arora et al. [27] used a convolutional neural network combined with a u-net [28], which can directly provide the pressure distribution around the airfoil without being adversely affected by manual data analysis. Parashar et al. [29] used deep learning to construct the flow field pressure from a Hessian tensor. Xie et al. [30] used heat flux as the target and modelled the sheet flux using a black box. Advantages over the traditional model have also been observed in the simple modelling of subgrid stresses [31]. Such research can avoid the uncertainty caused by the introduction of assumptions. Compared to the traditional method of theoretical analysis that combines various assumptions, such research also demonstrate a certain generalization performance, but generalization is difficult to guarantee.

To ensure the generalisation performance of the neural network model [32], a series of model training and application methods combined with discrete partial differential equation (PDE) have been developed. For example, Duraisamy [33] introduced a deep learning model by determining the mathematical form of the function, which provided better accuracy than the original model. Pan et al. [34] observed a good generalisation performance for a verification set using the first-order time discretisation of the PDE. Rassi [35] obtained the discrete form of the Navier-Stokes (N–S) equation from the perspective of experimental data using physics-informed neural networks (PINN) technology [36–38]. Wang [39] modelled the Reynolds stress error calculated by the RANS model from the perspective of the average flow field, with the help of physical constraints. This type of research can provide stable and reliable data-driven modelling results.

### B. Motivation, goals, and vision

The core of RANS modelling based on the eddy viscosity hypothesis is to obtain the eddy viscosity coefficient $\mu_t$. Therefore, additional hypotheses were introduced. For the two-equation model, the $k$ equations were derived strictly from the N–S equations, which is suitable for all flow fields where N–S equations are applicable. However, in the second equation, whether it is a $\varepsilon$ or $\omega$ equation, the respective assumptions introduced limited its applicability. The introduction of deep learning technology with physical restraints has presented the possibility of overcoming such limitations, thus, this study presents the phenomenological construction of a $k - \varepsilon$ equation source term based on the $\varepsilon$ model and broadens the applicability of the model.

During this study, although deep learning $\varepsilon$ modelling could effectively improve the accuracy of the simulation results in the flow field, the results after convergence are still not sufficiently accurate, and may even result in the failure of the constant simulation to converge to a definite solution (see Section II for details). This study argues that this failure is a result of the diverse physical phenomena, which provide a challenge for the basic assumptions of the viscous vorticity RANS model, causing the deviation in the iterative end points of the simulation from the training data, and the robustness of the phenomenological model is compromised. To solve this issue, a new technology (called "coordinate" in the text) was developed to generate a dataset for training a deep learning model, by modifying the $k$ field and $\varepsilon$ field that is then used for training corresponding to the real flow field, which can bring the simulation iteration endpoint of the constructed new model closer to the training data.

The remainder of this paper is organised as follows. In Section II, the construction method of the new source term and the training method of the neural network model are described, and the results of simulations utilising the new source term are presented. In Section III, the principle of the "coordinate" technology and the modification of the coordinate technology to the training data are presented. The resulting neural network model trained using the modified training data is also presented, which will be used to obtian a result in the numerical simulation. In Section IV, the performance of the numerical simulation of the new source term is summarized, and a comparison is made on whether or not to use the harmonic technology.

### II. New source items by deep learning.

#### A. Models and training methods

In the standard $k - \varepsilon$ model, the $\varepsilon$ equation is as shown in equation (1), where $\frac{\varepsilon}{k} C_{1\varepsilon}(G_k + C_{3\varepsilon}G_b)$ is the generation term. In this study, $C_{1\varepsilon}(G_k + C_{3\varepsilon}G_b)$ is replaced by the deep learning model $S_{DL} = f(\mu_t, S_{ij}, w)$, where $S_{ij}$ is the strain rate and $w$ is the internal parameter of the deep learning model. Therefore, the $\varepsilon$ equation becomes equation (2). Subsequently, the equation was discretised in time to obtain equation (3), where $\varepsilon^*$ is the dissipation rate of the next iteration step in CFD and $S_{DL}$ is the output by the neural network model. After time-averaging, the other variables on the right side of equation (3) and the input variables $\mu_t$ and $S_{ij}$ of the neural network were derived according to equations (4-6), so as to gnerate the training data set. Because this paper only discusses the modelling method, the real flow field data were replaced by the LES simulation data.

$$\frac{\partial(\rho\varepsilon)}{\partial t} + \frac{\partial(\rho\varepsilon u_i)}{\partial x_i} = \frac{\partial}{\partial x_j}\left[\left(\mu + \frac{\mu_t}{\sigma_\varepsilon}\right)\frac{\partial \varepsilon}{\partial x_j}\right]$$
$$+ \frac{\varepsilon}{k}C_{1\varepsilon}(G_k + C_{3\varepsilon}G_b) - C_{2\varepsilon}\frac{\rho\varepsilon^2}{k} \quad (1)$$

$$\frac{\partial(\rho\varepsilon)}{\partial t} + \frac{\partial(\rho\varepsilon u_i)}{\partial x_i} = \frac{\partial}{\partial x_j}\left[\left(\mu + \frac{\mu_t}{\sigma_\varepsilon}\right)\frac{\partial \varepsilon}{\partial x_j}\right]$$
$$+ \frac{\varepsilon}{k}S_{DL} - C_{2\varepsilon}\frac{\rho\varepsilon^2}{k} \quad (2)$$

$$\varepsilon^* = \Delta t\left\{-\frac{\partial(\varepsilon u_i)}{\partial x_i} + \frac{\partial}{\rho\partial x_j}\left[\left(\mu + \frac{\mu_t}{\sigma_\varepsilon}\right)\frac{\partial \varepsilon}{\partial x_j}\right] + \frac{\varepsilon}{\rho k}S_{DL} - C_{2\varepsilon}\frac{\varepsilon^2}{k}\right\}$$
$$+ \varepsilon \quad (3)$$

Each quantity is defined as follows:

$$k = \frac{1}{2}\sum \overline{u_i'^2} \quad (4)$$

$$\varepsilon = \frac{\mu}{\rho}\overline{\frac{\partial u_i'}{\partial x_j}\frac{\partial u_i'}{\partial x_j}} \quad (5)$$

$$\mu_t = \rho C_\mu \frac{k^2}{\varepsilon} \quad (6)$$

For a convergent steady solution, for any $\Delta t$, there should be $\varepsilon^* = \varepsilon$. Therefore, the neural network model is optimised as shown in equation (7):

$$\min_w |\varepsilon^* - \varepsilon| \quad (7)$$

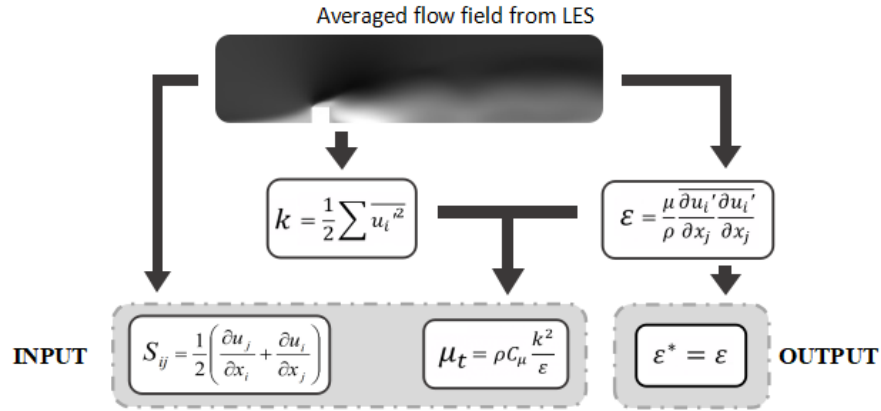

Figure 1  Process of obtaining feature quantities according to the original definition INPUT and OUTPUT for the deep learning model in this paper, which guides the data-driven training of the deep learning model.

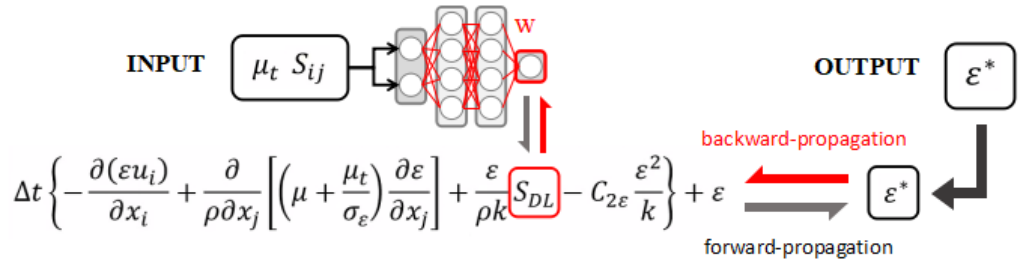

Figure 2  Methods for training deep learning models in this paper. Input and output were both derived from Figure 1, and the neural network model was only used as a schematic. In forward-propagation, the output result of INPUT after neural network $S_{DL}$ processing becomes a source term of the $\varepsilon$ equation, and the OUTPUT after the discretization time advancement was obtained. Training process is based on the PDE of discretization with the benefit that numerical stability is guaranteed as long as training is successful.

After the process of Figure 2, $w$ for the neural network model was obtained. As with convolutional neural networks, this neural network was trained with translational invariance for generalisation. This study focus on improving the target space of data-driven technology but not the improvement of accuracy caused by adjusting the number of layers and nodes of the neural network.  In order to minimize the influence of neural network details on the results, the deep learning model used LeakyRelu [40] as the activation function, which is an improvement of Relu [41], thereby preventing node necrosis. Use Kaiming initialization method [42] to initialize weights and biases. The number of nodes in each layer met the N + 4 criterion [43], reaching 32 neurones at the widest point. To eliminate the effect caused by the number of layers [44], this study draws on the residual network (ResNet) technology [45] to add two residual modules at the widest point. In addition, all gradient descents in this study were based on back-propagation [46], accelerated using the Adam [47] technology.

## B. Performance of the models

For the deep learning model are only trained by the data of the flow field after convergence, the CFD iterative process is not trained. Therefore, to guarantee numerical stability during RANS simulation, constraints must be imposed on the source terms of the deep learning construct: $0.7 \cdot S_{origin} < S_{ML} < 1.43 \cdot S_{origin}$, where $S_{origin}$ is the $C_{1\varepsilon}(G_k + C_{3\varepsilon}G_b)$ calculated when the new source term is replaced by the original $\varepsilon$ equation.

The data used in this study were derived from two-dimensional LES simulation results [48]. There is an absence of three-dimensional fully developed turbulence in the two-dimensional flow field, which is convenient for verifying the adaptability of the coordinate technology under extreme working conditions [49]. The LES simulation adopts a second-order upwind difference format, using a global isometric grid, with the first layer of grid y falling at 0.39 with a maximum Coulomb number of 0.039. The following adjudication was used to judge whether there was an average agreement.

$$\frac{L1Loss(u^n, u^{2n})}{L1Loss(u^{2n}, 0)} < 1\% \tag{8}$$

where $u^n$ and $u^{2n}$ are the mainstream directional average velocities at the nth and 2nth time step, and l1loss is the L1 norm of the two tensors.

In this study, the multiple average flow fields were divided into two parts: the training set and the validation set. The training set was constructed using three inlet flow fields with Reynolds numbers of 10,000, and their geometric channels were basically constructed using two-dimensional influx straight channels with an inlet width of 0.08 m and length of 0.4 m, with the different locations and sizes of squares set inside the channels as the trip lines. The validation set was not involved in training and was only used to detect the generalisation performance of the model, which was constructed from two flow fields with inlet Reynolds numbers of 10,000, which had different geometrical channels from the training set.

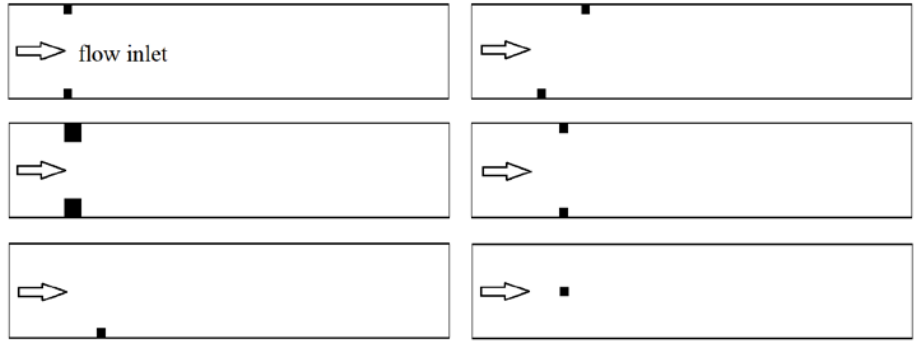

Figure 3 Two-dimensional flow field geometry channel schematic used in this study: Black sections represent the wall surface and internal solid; each channel up and down is the wall surface; left side is the fluid inlet and the right is the outlet.

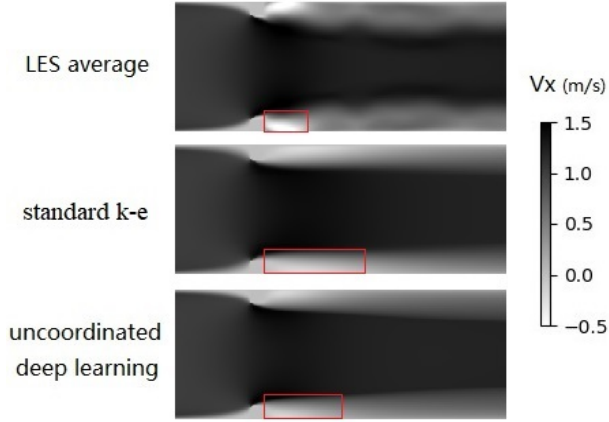

Figure 4  Deep learning model trained using the method described in Section II. Model successfully improves the prediction accuracy of the $k - \varepsilon$ model, and the boosting is most significant for the area of backflow behind the trip line, where the backflow zone is relatively shorter than the standard $k - \varepsilon$, with more drastic speed changes that are closer to the LES average. However, there are still some gaps, and the flow field state at the iterative convergence is intermediate between the standard $k - \varepsilon$ model and the training data (LES average outcome), still unmatched the training status.

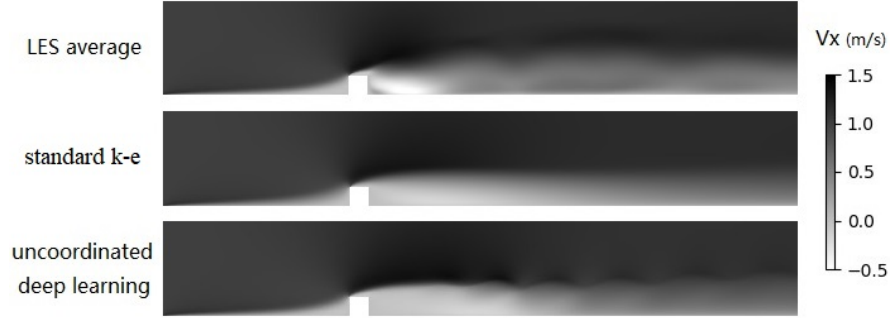

Figure 5  Training of a deep learning model using the method described in Section II for a particular flow field. This will cause the convergence to deteriorate and result in in the failure to obtain a constant result using normal solutions. This figure demonstrates the state in which the new model does not iteratively converge, the fluctuation exhibits periodicity that cannot be eliminated by continuing iteration, and the flow field cannot match the training state.

As shown in Figure 4 and Figure 5, using conventional deep learning training ideas can effectively improve the accuracy of flow field prediction, but there are still gaps in simulation, sometimes leading to convergence deterioration, which is largely caused by the mismatch between the physical assumptions of turbulence models and the real flow field. Such mismatches cause the flow field training data to not coincide with the simulation iteration endpoint of the model, and the iteration endpoint is far from the training state, exacerbating the deterioration of robustness.

## III. COORDINATE TECHNOLOGY

To improve the generality of the model to more varied flow field, this study proposes a "coordinate" technology. This technology will be used to generate the dataset for the training of the neural network, allowing the $\mu_t$ field to vary freely while automatically obtaining the optimal $k$ field and $\varepsilon$ field. The process is constrained by the $k$ equation and $\varepsilon$ definition as physical constraints, and is not restricted by other assumptions, so as to ensure the adaptability to changing flow conditions.

As shown in Figure 6, in this technology, the averaged sampled non-constant flow field is first acquired, then obtain the vortex adhesion field $\mu_t$ (III. A) in combination with manual interventions, after which the $k$ field with the $\varepsilon$ field (III. B) is acquired using the coordinate technology. Thus, the fields $\mu_t$, $k$, $\varepsilon$, and averaged velocity $u$ satisfy the Reynolds-Averaged equation to a degree in order to support the construction and modification of the phenomenological RANS turbulence model in the form of $k - \varepsilon$.

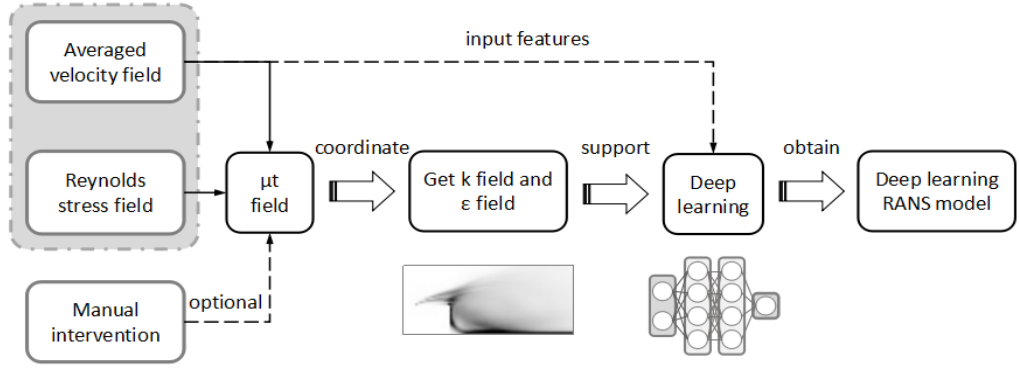

Figure 6 Coordinate method supports the process of RANS modelling. Core of this technology lies in generating the data sets required to train the neural network, without changing the iterative solution process when numerically simulated.

### A. Obtaining the $\mu_t$ field

The use of the coordinate method requires acquiring the $\mu_t$ field first. Considering the practically existing anisotropy of the real turbulent flow, it is not feasible to back-calculate the $\mu_t$ field according to the vortex stickiness assumption. Therefore, in this study, the optimal $\mu_t$ field is obtained using a data-driven optimisation method with the following principle:

In uncompressible flow, the Reynolds stress can be expressed as:

$$-\rho \overline{u_i u_j} = -\rho k \delta_{i,j} + \mu_t \left( \frac{\partial u_i}{\partial x_j} + \frac{\partial u_j}{\partial x_i} \right) \quad (9)$$

Then:

$$\overline{u_i u_j} = k \delta_{i,j} - \frac{\mu_t}{\rho} \left( \frac{\partial u_i}{\partial x_j} + \frac{\partial u_j}{\partial x_i} \right) \quad (10)$$

U, $\rho, \mu_t$, and $k$ can be obtained by sampling. Therefore, for any point in space, Equation (10) can be written as

$$\overline{u_i u_j} = f_{ij}(\mu_t) \quad (11)$$

The $\overline{u_i u_j}$ obtained by the average sampling process is denoted as $\overline{u_i u_j}^{\text{target}}$. Any one of the continuously loss functions $Loss(\text{output}, \text{target})$, denoted as $Loss_a$, can be optimised by gradient descent as follows:

$$\min_{\mu_t} \sum_{i=1}^{2} \sum_{j=1}^{2} Loss_a \left( f_{ij}(\mu_t), \overline{u_i u_j}^{\text{target}} \right) \quad (12)$$

The result of the back-calculation for the vortex viscous field $\mu_t$ can be regulated indirectly using the artificially obtained loss function $Loss_a$, which includes the weights in position versus the direction. In this study, the loss function $Loss_a$ is set to L1 form (emphasising that the form of the loss function is not unique and can be adjusted to demand), then equation (12) becomes the specific form as follows:

$$\min_{\mu_t} \sum_{i=1}^{2} \sum_{j=1}^{2} \left| f_{ij}(\mu_t) - \overline{u_i u_j}^{\text{target}} \right| \quad (13)$$

To ensure numerical legitimacy, after optimisation has been done to obtain the $\mu_t$ field, the minimum value among results needs to be artificially modified to $10^{-7}$ (far less than the physical viscosity).

### B. Obtaining the k field and ε field

When applying in engineering, the goal of RANS turbulence model is to obtain steady solutions through simulations. For any $k - \varepsilon$ model, the corresponding $k$ field versus $\varepsilon$ field during this time does not continue to change, and the local term between $k$ and $\varepsilon$ equations should be zero. However, the above discussion is conducted under ideal conditions. In the actual modelling process, the introduction of isotropic assumptions, which introduce inevitable errors, imbalances the equations. When using the model for numerical simulation the local term will no longer be zero. This means that the averaged field corresponding to the equation changes again and reduce the accuracy of the simulation results. The principle of the coordinate technology is to reduce the deviation of averaged flow field caused by error, automatically extrapolating the most stable $k$ field and the $\varepsilon$ field, and thus obtaining a further increase in accuracy. The specific method of this technology is given below.

In a flow field that is incompressible and without heat exchange, temperature vs density change, and reaches a steady state, the $k - \varepsilon$ model equation can be written as follows:

$$\frac{\partial (\rho k u_i)}{\partial x_i} = \frac{\partial}{\partial x_j} \left[ \left( \mu + \frac{\mu_t}{\sigma_k} \right) \frac{\partial k}{\partial x_j} \right] + 2\mu_t S_{ij} S_{ij} - \rho \varepsilon \quad (14)$$

$$\frac{\partial (\rho \varepsilon u_i)}{\partial x_i} = \frac{\partial}{\partial x_j} \left[ \left( \mu + \frac{\mu_t}{\sigma_\varepsilon} \right) \frac{\partial \varepsilon}{\partial x_j} \right]$$
$$+ \frac{\varepsilon}{k} \left( 2C_{1\varepsilon} \mu_t S_{ij} S_{ij} - C_{2\varepsilon} \rho \varepsilon \right) \quad (15)$$

When equation (6) exists, equation (15) can also be written as:

$$\frac{\partial(\rho k u_i)}{\partial x_i} = \frac{\partial}{\partial x_j}\left[\left(\mu + \frac{\mu_t}{\sigma_k}\right)\frac{\partial k}{\partial x_j}\right]$$
$$+ 2\mu_t S_{ij} S_{ij} - \rho^2 C_\mu \frac{k^2}{\mu_t} \quad (16)$$

where, the coefficient $C_\mu$ is 0.09 and $\sigma_k$ is 1.

$\varepsilon$ has been eliminated in equation (16). When the average velocity field, numerical format, and $\mu_t$ field are determined using the three ways mentioned, the $k$ field is the only unknown in equation (16). At this point, the optimal $k$ field solution can be obtained using the gradient descent optimisation method.

The actual optimisation is performed based on discretisation, and the local term after discretisation is expressed as:
$$\frac{\partial(\rho k)}{\partial t} = \frac{\rho(k^* - k)}{\Delta t} \quad (17)$$

Equation (16) becomes:
$$k^* = k + \frac{\Delta t}{\rho} IT(k) \quad (18)$$

Where:
$$IT(k) = -\frac{\partial(\rho k u_i)}{\partial x_i} + \frac{\partial}{\partial x_j}\left[\left(\mu + \frac{\mu_t}{\sigma_k}\right)\frac{\partial k}{\partial x_j}\right]$$
$$+ 2\mu_t S_{ij} S_{ij} - \rho^2 C_\mu \frac{k^2}{\mu_t} \quad (19)$$

For any point in space, all but the $k$ in the formula are known. Equation (18) can therefore be written as
$$k^* = f(k) \quad (20)$$

While performing the optimization, the loss function takes the L1 form:
$$\min_k |f(k) - k| \quad (21)$$

Optimization of Equation (21) by means of gradient descent allows the $k$ field to approach a constant state. After obtaining the $k$ field, the $\varepsilon$ field was calculated directly using Equation (6).

A new vocabulary has been introduced in this paper for the ease of presentation. The target fields obtained using Equations (21) and (6) are called the 'coordinated' turbulent kinetic energy field $k$ and the turbulent dissipation rate field $\varepsilon$. The deep learning model trained on the basis of these fields is called the 'coordinate' deep learning model. Corresponding fields obtained using equations (4), (5), and (6) with the deep learning model were all described as 'uncoordinated'.

## IV. RESULTS AND DISCUSSIONS

### A. Coordinate result is affected by the flow state

The coordinate technology, which automatically modifies the target $k$ field versus the $\varepsilon$ field according to the actual flow state should have the following properties: when the actual flow state complies with the existing turbulence model assumptions, the modification results should be degraded to those calculated by this turbulence model, and as the flow regime progressively deviates from the assumptions of this turbulence model, the gap in the calculated results for the modified $k$ field versus the $\varepsilon$ field in this model becomes more apparent. This part of the study is therefore used to examine whether a coordinate technology possesses this property. The simulation results, using the standard $k - \varepsilon$ model as a known turbulence model, were compared to the $k$ field constructed from the coordinate technology.

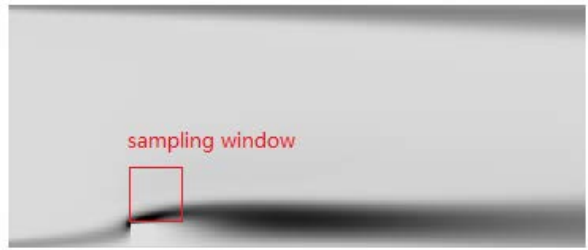

Figure 7 Plot of the target $k$ field acquired for the standard $k - \varepsilon$ model, where the red box is the sampling window. Test data were sampled within the window only, independent of import/export boundary conditions.

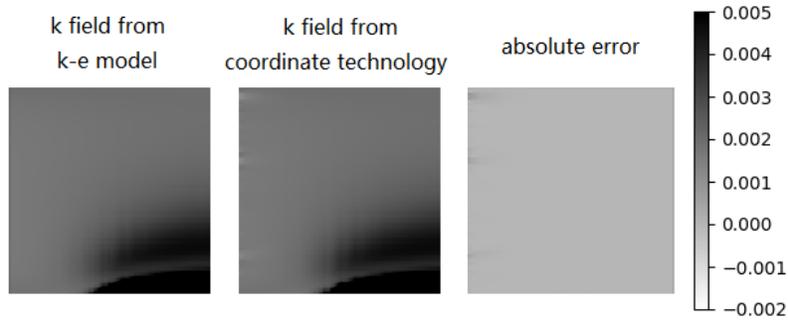

Figure 8 Upper panel shows the validation results of the coordinate technology. Absolute value error for the $k$ field obtained using the modification of the coordinate technology fits perfectly with the original $k$ field, except for the minor imperfections at the boundaries, in accordance with the expectations in this paper. This result indicated that the target $k$ field modified by the coordinate technology was consistent with the optimal $k$ field, and such agreement was largely unaffected by the boundary conditions.

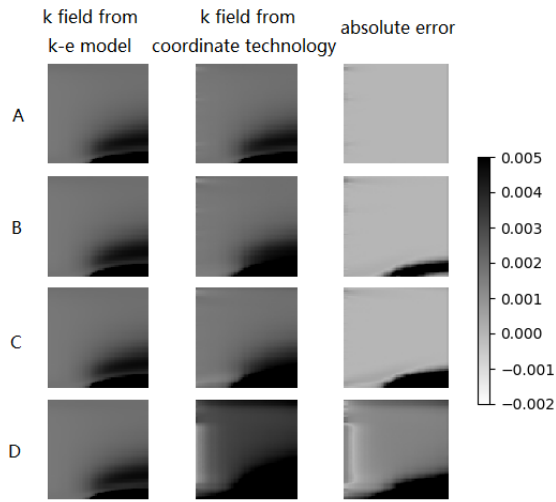

Figure 9 Illustration of the effects of actual versus assumed inconsistencies on the optimal $k$ field, which is achieved using a wrong $\mu_t$ field. Where group A was not modified, the minimum value in the restricted $\mu_t$ field was 1e-4 in group B, 3e-4 in group C, and 1e-3 in group D. Modified $k$ field changes with the measured $\mu_t$ field. Higher the degree to which the $\mu_t$ field deviates from the computed value of the turbulence model, the higher the degree to which the $k$ field deviates from the truth.

Figure 9 shows the tests when the actual flow field is consistent with the assumptions of the turbulence model. The $k$ field given by the coordinate technology should be consistent with the $k$ field calculated by the turbulence model. When the physical reality deviates from the assumptions, such as the isotropic assumption or the complete turbulence assumption, the modification results should be different from the model-calculated $k$ field. The optimal $k$ field constantly changes with the introduction of the actual deviation from the hypothesis.

However, the grooming ability of the coordinate technology decreases as the $\mu_t$ field deviates from the turbulence model, and the convergence speed and the limiting convergence accuracy of the coordinate technology deteriorates gradually. This is because when the flow regime deviates from the standard assumptions, the coordinate technology is used to adjust the spatial distribution of the transport volume by adjusting the velocity field, which serves as the coordinate benchmark and does not follow the change in the velocity field, the k field, and the ε field, which cannot precisely match the transport equation. The more severe the deviation in the flow regime, the more difficult it is to satisfy the transport equation. Thus, the ability of coordinate technology to correct model assumptions is limited.

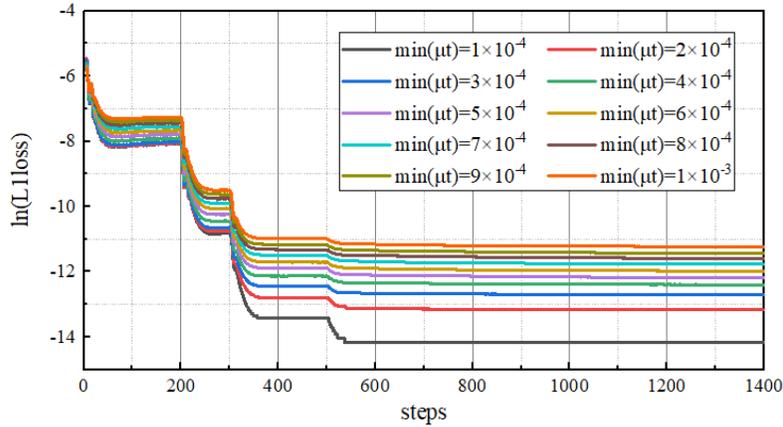

Figure 10  Graph showing the decline in the L1 Loss in Equation 20. Lower the L1 loss value, the higher the match of the $k$ field to the $k$ equation, and the closer the corresponding $k$ field is to the norm. $\mu_t$ field related to each curve in the plot corresponds to different artificial errors, and the given method of error is shown in Figure 9 by min ($\mu_t$). Larger the target $\mu_t$ field error, the worse the decrease in the L1 loss, the more the $k$ field deviates from the steady solution

## B. Accuracy improvement by coordinate

According to Section III of this paper, the boosting effect of coordinate technology on the deep learning RANS turbulence modelling was tested. For the convenience of alignment, LES sampling results, standard $k - \varepsilon$ model simulation results, uncoordinated simulation results, and reconciled simulation results were presented at the same time. LES sampling results were used to train the model while serving as a benchmark for evaluation. The performance of the coordinate-technology-based RANS model on the more typical training and validation sets is given in detail here.

In the training set example below, the introduction of the coordinate technology reduces the prediction error for the reattachment site locations by 38.5 %. For the average flow field, the coordinate technology reduces the error by 6.2 % based on the deep learning method and by 51.7 % for the standard $k - \varepsilon$ model. This result demonstrates that modifying the training data using coordinate technology can effectively improve the accuracy of the simulation results.

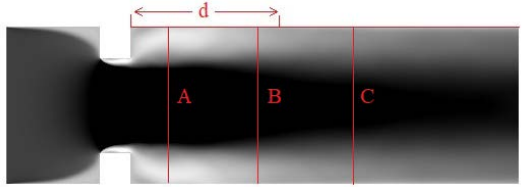

Figure 11  Arrangement of the three sampling lines (A, B, C) for the set of training data in the example set. Velocity near the wall at $y^+ = 0.39$ was detected after the tripping line.

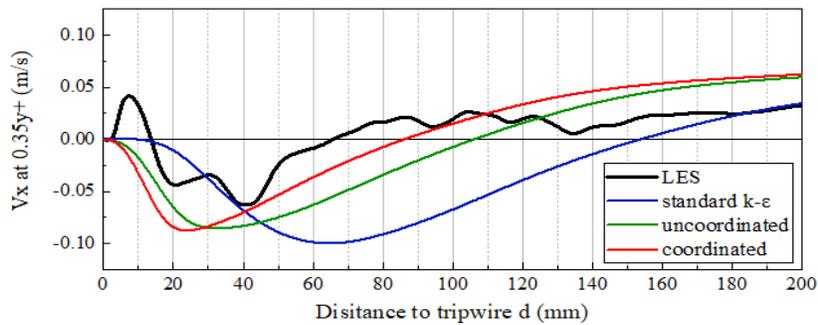

Figure 12  Improvement of the prediction of the reattachment point using the coordinate technology. Target reattachment point was at d = 65.8 mm, which was located at d = 153.8 mm using the standard $k - \varepsilon$ model, and the largest reattachment point was at d = 106.1 mm using the uncoordinated conventional machine learning method, which gave better prediction than the standard $k - \varepsilon$ model. Coordinate machine learning method shows that the location of this point is at d = 86.1 mm, which further improves the accuracy of the reattachment point and reduces the simulation error by 38.5 %.

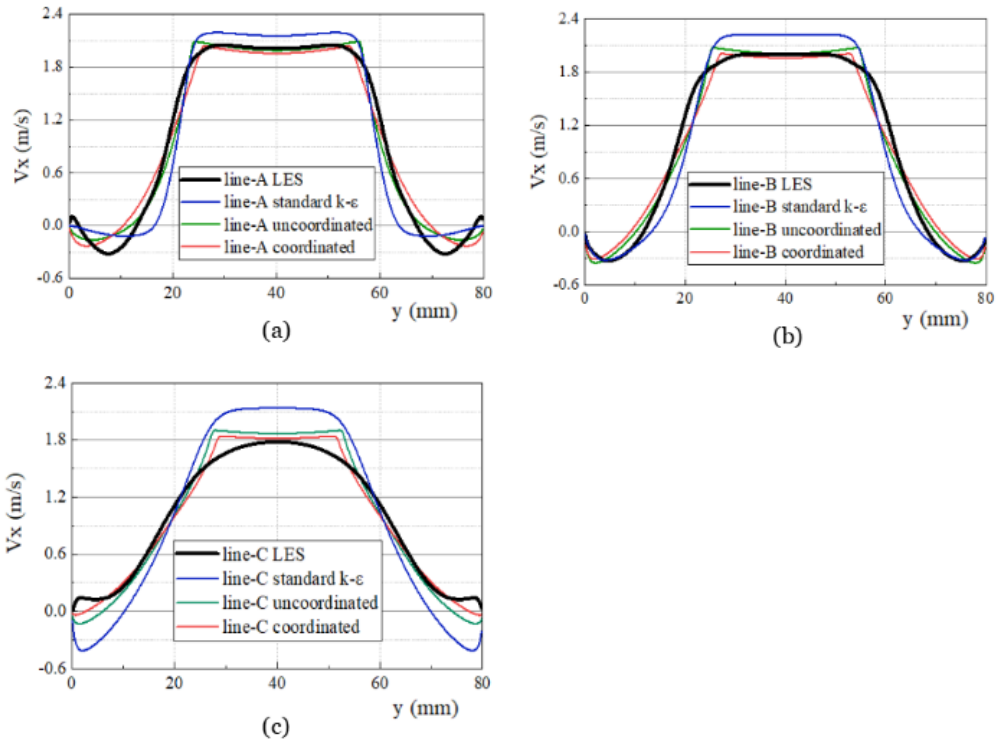

Figure 13 Improved effects of coordinate technology on the simulation results of the overall flow field. From the velocity type profiles, the more downstream the location is, the more advantageous the coordinate technology is. Deep learning technology significantly improves the flow distribution in the mainstream region, and the coordinate technology further improves the accuracy, and finally make the simulation results to be the closest to the target flow field obtained by LES.

The coordinate technology showed better generalisation performance for the validation set, giving better prediction results for the location of the reattachment point. The error of the whole field decreased by 19.5 % averagely before applying coordinate technology compared to 29.5 % after applying. This result indicates that the coordinate technology does not reduce the generalisation of traditional deep-learning methods.

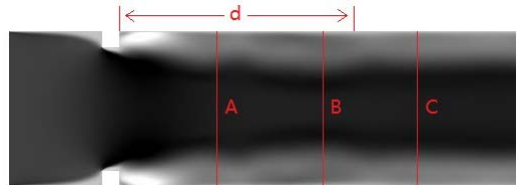

Figure 14 Averaged LES results as the validation set, where the red line indicates the data sampling location. Sampling lines were arranged at three spatial location A, B, and C.

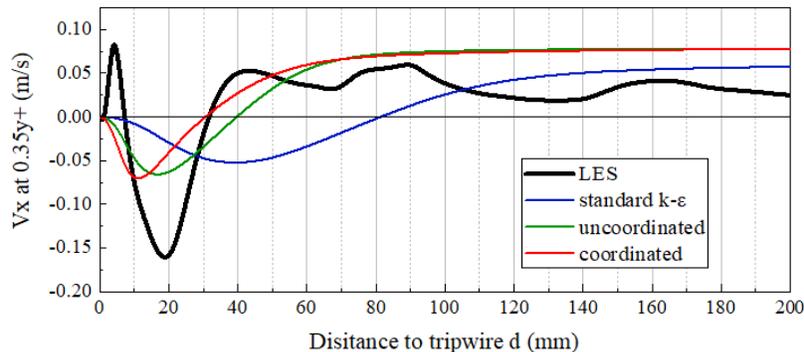

Figure 15 Improvement in predicting reattachment point using the coordinate technology. Target flow field reattachment point position was at d = 31.7 mm; d = 81.4 mm for the standard $k - \varepsilon$ model; d = 39.7 mm for the uncoordinated deep learning model, and d = 30.6 mm for the coordinate machine learning model. Coordinate reduces the simulation error by 86.3 %.

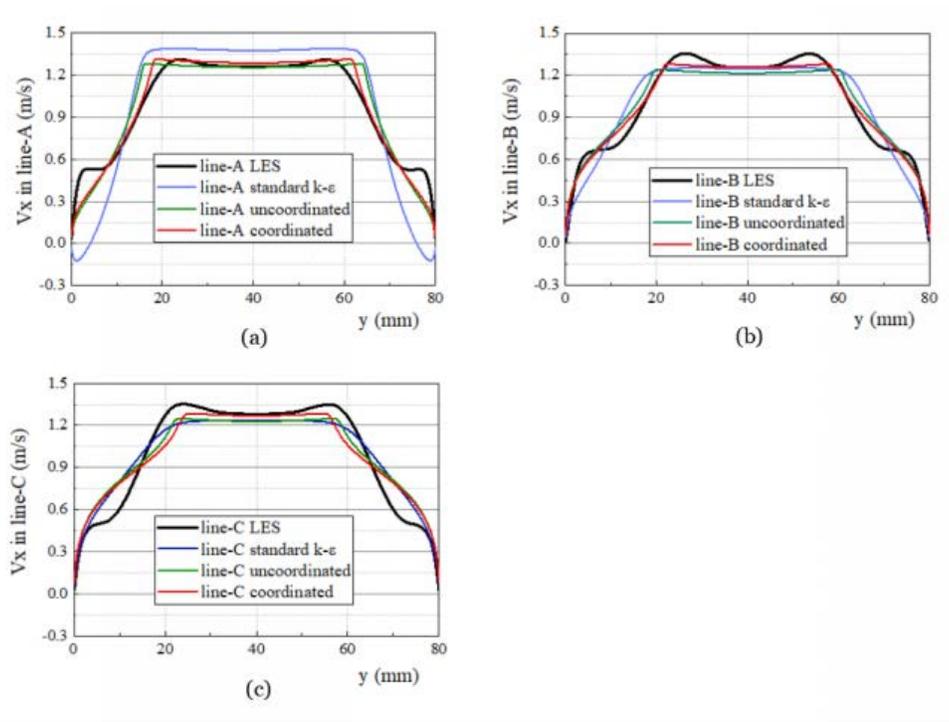

Figure 16  X-direction flow rate data obtained on the sampling location. In Figure 16(a), the prediction results for the standard $k - \varepsilon$ model shows that the wall flow is separated, which is conflict with the results given by LES and indirectly leads to a much worse rate of flow distribution, which causes the main flow velocity to be significantly higher than it should be. After using the coordinate technology, the simulation results gave the closest velocity distribution to the target flow field near the wall surface, thus bringing a better rate of flow distribution in the mainstream area and significantly improving the prediction results of the main stream velocity.

### C. Convergence performance improvement by coordinate technology

In addition to accuracy improvement, the coordinate technology improve the convergence performance when approaching a steady solution , which means that the coordinate method is able to alleviate the bad convergence performance that arises in the RANS turbulence modelling to a certain degree. The use of the traditional deep learning construction scheme for the flow field shown in Figure 17 introduces non-constancy (II. B), and instead poses more obstacles for numerical simulation. After the coordinate technology was introduced, the iterative convergence state of the flow field was closer to the original sampling data state, and the non-constancy was suppressed.

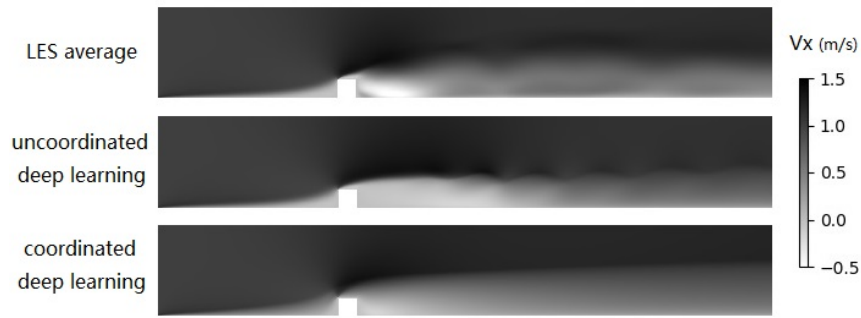

Figure 17  Modelling of the uncoordinated deep learning RANS deteriorates in convergence during practical use, and there are fluctuations in the trailing points corresponding to stumbling lines. After the coordinate technology was introduced, the constant convergence of the simulation results obtained demonstrated a clear improvement, and the fluctuation in the wake disappeared.

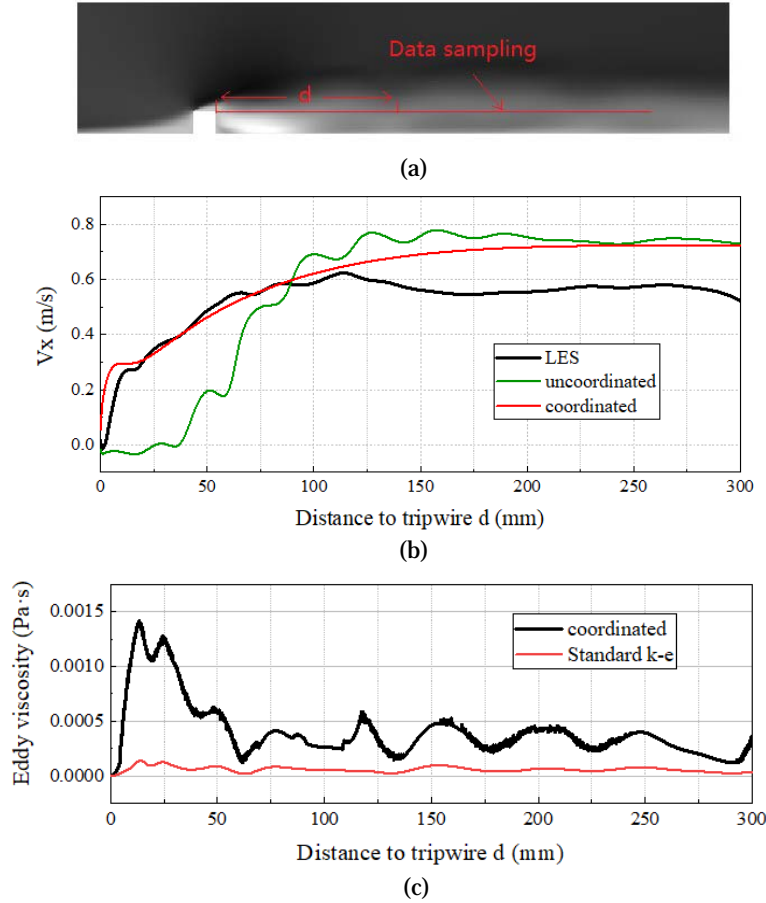

(a)

(b)

(c)

Figure 18 Sampling data of the corresponding flow field. Figure 18 (b) is the x-direction velocity along the sampling line, where the unadjusted model exhibits periodic fluctuations for constant solutions. In Figure 18 (c), the local $\mu_t$ values were extracted and used to obtain the vortex viscosity fields calculated according to Equations (4-6), which were significantly different from those calculated using Equation (13), and the level of vortex viscosity obtained using Equation (13) was substantially higher than the estimated values from conventional calculation methods.

For the convergence enhancement phenomenon brought about by the introduction of the coordinate method, this study uses a qualitative method to analyze the reasons. Data sampling was performed following the trip line location shown in Figure 18 (a), and the X-direction speed values were extracted separately from the local $\mu_t$ values and presented in 18 (b) and 18 (c). The low levels of the target $\mu_t$ field values are the source of the problem, which means that data driving using the training data acquired via the standard method introduces a large bias in cases where the flow regime does not satisfy the assumptions. In conventional schemes, this occurs when the targeted theoretical corrections are made, increasing the number of transport equations and discarding the isotropic hypothesis. However, after the coordinate technology was introduced, the algorithm was able to adapt to the special working conditions automatically even without theoretical modification, guaranteeing the accuracy of the simulation.

## V. CONCLUSIONS

In this study, using deep learning technology, we phenomenologically constructed a new source term in place of the $\varepsilon$ equation generation term. Simultaneously, to improve the general applicability of the model to various flow fields, a "coordinate" technology was developed to make the automatic creation of the neural network training process more consistent with the theoretical assumptions and further improve the simulation accuracy of the deep learning RANS turbulence model. This study tested the characteristics of the coordinate technology using numerical experiments and aligned the simulation results on multiple training sets with the validation set, drawing the following conclusions:

1、For the simulation results for the standard $k - \varepsilon$ model, the introduction of new source terms effectively reduces the whole field average error for the simulation results but still cannot fully match the target flow field, with a hidden risk of convergence deterioration.

2、Coordinate technology can adapt the modified $k$ field to the $\varepsilon$ field training data. When the real flow state deviates from the assumption, the modified $k$ field also produces a synchronous shift from the $\varepsilon$ field, thus bringing the training data as close as possible to the final convergence state of the control equation. The addition of coordinate technology can further improve the simulation accuracy of the data-driven RANS turbulence model with iterative convergence.

3、The data-driven RANS turbulence model obtained by coordinate technology is more robust and can be effectively generalised with relatively similar flows; with changing the size and location of geometric features not substantially affecting the simulation accuracy.

4、The training using the $\varepsilon$ field, $k$ field, and $\mu_t$ field derived from the definition is not effective; a possible reason being that $\bar{u}, \varepsilon, k,$ and $\mu_t$ of the real flow field in the complex case cannot meet the vortex stickiness assumption. The turbulent vortex mass stickiness coefficient formula using the $k$ equation, which in turn causes the flow field state corresponding to the iterative end point of the simulation model to deviate from the training state. The simulation at this point relies on the generalisation performance of the model, and the robustness is relatively deteriorated. The training data after modification with coordinate technology is closer to the iteratively converged state of the numerical simulation, making the trained model have a less generalizable uncertainty. The numerical experiments in this study also show that using the "coordinate" technology to train the resulting model increases the accuracy of numerical simulation results.

## COMPETING INTERESTS

The authors do not have any conflicts of interest to declare.

## FUNDING

This work was supported by the National Natural Science Foundation of China [grant number 51906008]; Fundamental Research Funds for the Central Universities [grant number YWF-21-BJ-J-822]; and National Science and Technology Major Project [grant number 2017-Ⅲ-0003-0027]. The funding sources were not involved in study design; in the collection, analysis and interpretation of data; in the writing of the report; and in the decision to submit the article for publication.

## AUTHOR'S CONTRIBUTION

Shuming Zhang: Formal analysis, Investigation, Writing-original draft. Haiwang Li: Supervision, Conceptualization. Ruquan You: Conceptualization, Data curation Writing-review & editing. Tinglin Kong: Writing-review & editing. Zhi Tao: Supervision, Conceptualization.